\documentclass[showpacs,aps,graphicx]{revtex4}
\usepackage{amsmath}
\usepackage{graphicx}

\usepackage{multirow}

\begin{document}

\title{Quantum state transfer and controlled-phase gate on one-dimensional
superconducting resonators assisted by a quantum
bus\footnote{Published in Sci. Rep. \textbf{6}, 22037 (2016)}}

\author{Ming Hua, Ming-Jie Tao, and Fu-Guo Deng\footnote{Corresponding author: fgdeng@bnu.edu.cn}}

\address{Department of  Physics, Applied Optics Beijing Area Major Laboratory, Beijing
Normal University, Beijing 100875, China}

\date{\today }

\begin{abstract}
We propose a quantum processor for the scalable quantum computation
on microwave photons in distant one-dimensional superconducting
resonators. It is composed of a common resonator $R$ acting as a
quantum bus and some distant resonators $r_j$  coupled to the bus in
different positions assisted by superconducting quantum
interferometer devices (SQUID), different from  previous processors.
$R$ is coupled to one transmon qutrit, and the coupling strengths
between $r_j$ and $R$ can be fully tuned by the external flux
through the SQUID. To show the processor can be used to achieve
universal quantum computation effectively, we present a scheme to
complete the high-fidelity quantum state transfer between two
distant microwave-photon resonators and another one for the
high-fidelity controlled-phase gate on them. By using the technique
for catching and releasing the microwave photons from resonators,
our processor may play an important role in quantum communication as
well.
\end{abstract}

 \maketitle

\section{Introduction}

Quantum computation \cite{Nielsen}, which can implement the famous
Shor's algorithm \cite{shor} for integer factorization and
Grover/Long algorithm \cite{Grover,LongGrover}  for unsorted
database search,  has attracted much attention in recent years.
There are some interesting systems which have been used to realize
quantum computation, such as photons \cite{Knill,Ren2}, quantum dots
\cite{Hu1,WeiSR}, nuclear magnetic resonance \cite{NMR,Long1,Long2},
diamond nitrogen-vacancy center \cite{Togan,Neumann}, and cavity
quantum electrodynamics (QED) \cite{Nielsen}. Achieving quantum
computation, quantum state transfer \cite{Sillanpaa,Majer} and
universal quantum gates have been studied a lot, especially the
two-qubit controlled-phase (c-phase) gate or its equivalent
(controlled-not gate) which can be used to construct a universal
quantum computation assisted by single-qubit operations
\cite{Nielsen}. To construct the high-efficiency and high-fidelity
quantum state transfer and the c-phase gate on fields or atoms,
cavity QED, composed of a two-energy-level atom coupled to a
single-mode filed, has been studied a lot.

Simulating cavity QED, circuit QED
\cite{Blais,Wallraff,Tian,Miranowicz,Leghtas,Xue2,Sun,Stojan,Hu,Du,Yang1,Yang2},
composed of a superconducting qubit coupled to a superconducting
resonator, plays an important role in  quantum computation because
of its good ability for the large-scale integration
\cite{Chow2,Cao,You,Galiautdinov,Kelly,Hua4}. By far, some important
tasks of quantum computation based on the superconducting qubits
have been realized in experiments. For example, DiCarlo \emph{et
al.} demonstrated a c-phase gate on two transmon qubits
\cite{DiCarlo1} in 2009, and they prepared and measured the
entanglement on three qubits in a superconducting circuit
\cite{DiCarlo2} in 2010.  In 2012, Lucero \emph{et al.}
\cite{Lucero} computed the prime factors with a Josephson phase
qubit quantum processor and Reed \emph{et al.} \cite{Reed}
constructed a controlled-controlled phase gate to realize a
three-qubit quantum error correction with superconducting circuits.
In 2014, Barends \emph{et al.} \cite{Barends} realized the
single-qubit gate and the c-phase gate on adjacent Xmon qubits with
high fidelities of $99.94\%$ and $99.4\%$, respectively.

Interestingly, the quality factor of a one-dimensional (1D)
superconducting resonator \cite{Megrant} has been enhanced to
$10^6$, which makes the resonator as a good carrier for quantum
information processing
\cite{Sandberg,Merke,Liao,Strauch3,Eichler,Strauch4,Li,Chen,Liu,Yan,Hua1,Hua3,Nori}.
For instance, Houck \emph{et al.} \cite{Houck} generated single
microwave photons in a circuit in 2007. In 2008, Hofheinz \emph{et
al.} \cite{Hofheinz} generated the Fock states in a superconducting
quantum circuit. In 2010, Johnson \emph{et al.} \cite{Johnson}
realized the quantum non-demolition detection of single microwave
photons in a resonator. In 2011, Wang \emph{et al.} \cite{Wang}
deterministically generated the entanglement of photons in two
superconducting microwave resonators and Strauch \emph{et al.}
\cite{Strauch1} proposed a scheme to prepare the NOON state on two
resonators. In 2013, Yang \emph{et al.} \cite{Yang} presented two
schemes for generating the entanglement between microwave photons
and qubits. Recently, Hua \emph{et al.} \cite{Hua2} proposed some
schemes to construct the universal c-phase and cc-phase gates on
resonators.

There have been some theoretic studies on constructing the
multi-resonator quantum entanglement and the universal quantum gate
on local microwave-photon resonators in a processor composed of some
resonators coupled to a superconducting qubit
\cite{Strauch1,Yang,Hua2,Strauch2,Wu}. In this paper, we propose a
quantum processor for quantum computation on distant resonators with
the tunable coupling engineering \cite{Tsomokos,Peropadre} between
the superconducting resonator and the quantum bus.  There is just
one superconducting transmon qutrit $q$ in our processor,  which is
coupled to a common resonator $R$ (acts as a quantum bus). Different
from the processors in previous works
\cite{Strauch1,Yang,Hua2,Strauch2,Wu}, the resonators $r_j$
($j=1,2$) (act as the information carriers) in our processor are
coupled to the quantum bus $R$, not the qutrit, which makes it have
the capability of integrating some distant resonators
\cite{DiCarlo2} by coupling them to the bus in different positions.
In contrast with the resonator-zero-qubit architecture by
Galiautdinov \emph{et al.} \cite{Galiautdinov}, the resonators in
our processor are used for quantum information processing, not the
memory elements. It does not require more superconducting qubits.
With our processor, we present an effective scheme for the quantum
state transfer between $r_1$ and $r_2$ with the Fock states
$|0\rangle_j$ and $|1\rangle_j$ and another for the c-phase gate on
two resonators by using the resonance operations between $R$ and
$r_j$ and that between $R$ and $q$. The fidelities of our quantum
state transfer and c-phase gate are $99.97\%$ and $99.66\%$,
respectively. By catching and releasing the microwave photons from
resonators \cite{Yin}, our processor maybe play an important role in
quantum communication.

\begin{figure}[h]
\centering
\par
\includegraphics[width=8 cm,angle=0]{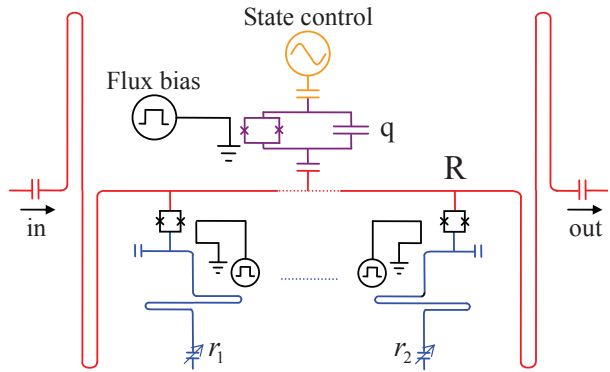}
\caption{(Color online)  Schematic diagram for the construction of
the quantum state transfer between the two microwave-photon
resonators $r_j$ ($j=1,2$) and the c-phase gate on $r_j$ assisted by
a quantum bus (i.e., the common resonator $R$) which is coupled to
only a superconducting transmon qutrit $q$.}
\end{figure}

\section{Results}

\subsection{Quantum processor composed of  resonators and a quantum bus}

Our quantum processor is composed of some distant high-quality 1D
superconducting resonators $r_j$ and  a high-quality 1D
superconducting resonator $R$, shown in Fig. 1. The common resonator
$R$ acts as a quantum bus for quantum information processing and it
is capacitively coupled to a $\Xi$ type three-energy-level
superconducting transmon qutrit $q$ whose frequency can be tuned by
an external magnetic field. The qutrit is placed at the maximum of
the voltage standing wave of $R$ (not be drawn here). The simple
superconducting quantum interferometer device (SQUID) with two
Josephson junctions inserted between $r_j$ and $R$ serves as the
tunable-coupling function between them. The SQUID variables are not
independent and introduce no new modes \cite{Peropadre}. Here, the
SQUIDs are not sensitive to the charge noise and can achieve a full
tunability. Besides, the plasma frequencies of SQUIDs should be
larger than the frequencies of the resonators. $r_{j}$  are  laid
far enough to each other to avoid their direct interaction generated
by mutual capacitances and mutual inductive coupling.  In the
interaction picture, the Hamiltonian of the processor is ($\hbar
=1$, under the rotating-wave approximation)
\begin{eqnarray}        
\begin{split}
H =& g_{g,e}\left(a_{_R}\sigma_{g,e}^{+}e^{i\Delta_{g,e}
t}+a_{_R}^{+}\sigma_{g,e}^{-}e^{-i\Delta_{g,e} t}\right)
+ g_{e,f}\left(a_{_R}\sigma_{e,f}^{+}e^{i\Delta_{e,f} t}+a_{_R}^{+}\sigma_{e,f}^{-}e^{-i\Delta_{e,f} t}\right) \\
& + \sum_{j=1,2} g_{j}\left(b_{j}^{+}a_{_R}e^{i\Delta_j t} +
b_{j}a_{_R}^{+}e^{-i\Delta_j t}\right). \label{Hall}
\end{split}
\end{eqnarray}
Here,  $\Delta_{g,e(e,f)}=\omega_{g,e(e,f)}-\omega_{_R}$ and
$\Delta_j=\omega_{j}-\omega_{_R}$. $\omega _{_R}$ and  $\omega _{j}$
are the the first mode frequencies of $R$ and $r_{j}$, respectively.
$\omega_{g,e(e,f)}$ is the frequency of the transmon qutrit $q$ with
the transition $|g\rangle \leftrightarrow |e\rangle$ ($|e\rangle
\leftrightarrow |f\rangle$) in which $|g\rangle$, $|e\rangle$, and
$|f\rangle$ are the ground, the first excited, and the second
excited states of the qutrit, respectively. $a_{_R}^{+}$ and $
b_{j}^{+}$ are the creation operators of $R$ and $r_{j}$,
respectively. $\sigma_{g,e}^{+}= |e\rangle \langle g|$ and
$\sigma_{e,f}^{+}= |f\rangle \langle e|$ are the creation operators
of the two transitions of $q$, respectively. $g_{g,e}$ and $g_{e,f}$
($g_{e,f}=\sqrt{2}\,g_{g,e}$) are the coupling strengths between $R$
and the two transitions of $q$, respectively. $g_{j}$ is the
coupling strength between $r_j$ and $R$, which is contributed by
their capacitive and inductive and can be tuned by the external flux
through the SQUID \cite{Peropadre}.

The evolution of our processor can be described by the master
equation \cite{ABlais}
\begin{eqnarray}              
\begin{split}
\frac{d\rho }{dt}  =&  -i[H,\rho ]+\kappa_1 D[b_1]\rho + \kappa_2
D[b_2]\rho + \kappa_R D[a]\rho
+ \gamma_{g,e}D[\sigma_{g,e}^{-}] \rho +\gamma_{e,f}D[\sigma_{e,f}^{-}]\rho   \\
& +
\gamma_{\phi,e}(\sigma_{ee}\rho\sigma_{ee}-\sigma_{ee}\rho/2-\rho\sigma_{ee}/2)
+
\gamma_{\phi,f}(\sigma_{ff}\rho\sigma_{ff}-\sigma_{ff}\rho/2-\rho\sigma_{ff}/2).
\label{masterequation}
\end{split}
\end{eqnarray}
Here, the operator $D[L]\rho =(2L\rho L^{+}-L^{+}L\rho -\rho
L^{+}L)/2$ ($L=a$, $b$, $ \sigma _{g,e}^{-}$, $\sigma _{e,f}^{-}$).
$\sigma_{ee}=|e\rangle\langle e|$ and $\sigma_{ff}=|f\rangle\langle
f|$.  $\kappa_1$, $\kappa_2$, and $\kappa_R$ are the decay rates of
the resonators $r_1$, $r_2$, and $R$, respectively.  $\gamma_{g,e}$
($\gamma_{e,f}$) is the energy relaxation rate of the qutrit with
the transition $|e\rangle \rightarrow |g\rangle$ ($|f\rangle
\rightarrow |e\rangle$). $\gamma_{\phi,e}$ ($\gamma_{\phi,f}$) is
the dephasing rate of the level $\vert e\rangle$ ($\vert f\rangle$)
of the qutrit. To achieve the resonance operations between $R$ and
$r_j$, the transition frequencies of all the resonators are taken
equal to each other.


\subsection{Quantum state transfer between $r_1$ and $r_2$}

Our quantum-state-transfer protocol between $r_1$ and $r_2$ can be
completed with two resonance operations between the quantum bus $R$
and the resonator $r_j$. The interaction between $R$ and $r_j$ can
be described as
\begin{eqnarray}           
H^{R,r_j} =g_{j}\left(b_{j}^{+}a_{_R}e^{i\Delta_j t} +
b_{j}a_{_R}^{+}e^{-i\Delta_j t}\right). \label{Rr}
\end{eqnarray}
In our scheme, the states $|0\rangle_{_R} |0\rangle_j$,
$|0\rangle_{_R} |1\rangle_j$, and $|1\rangle_{_R} |0\rangle_j$ are
required only. Here, the state $|0\rangle_{_R} |0\rangle_j$ keeps
unchanged with the evolution $U^{R,r_j}=e^{-i H^{R,r_j} t}$.
$|n_{_R}\rangle_{_R}$ and $|n_r\rangle_j$ are the Fock states of $R$
and $r_j$, respectively. $n_{_R}=a_{_R}^{+}a_{_R}$ and
$n_r=b_{j}^{+}b_{j}$. For the resonance condition between $R$ and
$r_j$ ($\Delta_j=0$) and if we take the initial state of the
subsystem composed of $R$ and $r_j$ to be $|0\rangle_{_R}
|1\rangle_j$, the state of the system composed of $R$ and $r_j$ can
be expressed as (further details can be found in the method)
\begin{eqnarray}           
|\psi(t)\rangle= \cos (g_j t)|0\rangle_R |1\rangle_j-i\sin (g_j
t)|1\rangle_R |0\rangle_j. \label{2}
\end{eqnarray}

Our scheme for the quantum state transfer between the two resonators
$r_1$ and $r_2$ can be accomplished with two-step resonance
operations described in detail as follows.

Initially, we assume the initial state of the processor is
\begin{eqnarray}           
|\psi\rangle_0^{transfer} = (\cos{\theta}|0\rangle_1 +
\sin{\theta}|1\rangle_1) \otimes |0\rangle_{_R} \otimes |0\rangle_2
\otimes |g\rangle, \label{transfer0}
\end{eqnarray}
 which means $r_1$ is in the state
$\cos{\theta}|0\rangle_1+\sin{\theta}|1\rangle_1$, $R$ and $r_2$ are
all in the vacuum state, and $q$ is in the ground state. First,
tuning the transition frequency of $q$ to detune with $R$ largely
and turning off (on) the coupling strength between $R$ and $r_2$
($r_1$) by using the external flux through their SQUIDs, the state
of the processor can evolve into
\begin{eqnarray}           
|\psi\rangle_1^{transfer} = |0\rangle_1 \otimes
(\cos{\theta}|0\rangle_{_R} - i\sin{\theta}|1\rangle_{_R}) \otimes
|0\rangle_2 \otimes |g\rangle \label{transfer1}
\end{eqnarray}
after a time of $g_1 t=\pi/2$.

Second, keeping the frequency of $q$ detune with $R$ largely,
turning off $g_1$, and turning on $g_2$, the state of the processor
can evolve from Eq. (\ref{transfer1}) to
\begin{eqnarray}           
|\psi\rangle_2^{transfer} = |0\rangle_1 \otimes |0\rangle_{_R}
\otimes (\cos{\theta}|0\rangle_2 - \sin{\theta}|1\rangle_2) \otimes
|g\rangle \label{transfer2}
\end{eqnarray}
within a time of $g_2 t=\pi/2$. Here, we complete the quantum state
transfer as
\begin{eqnarray}           
|0\rangle_1 \otimes (\cos{\theta}|0\rangle_{_R} +
\sin{\theta}|1\rangle_{_R}) \otimes |0\rangle_2 \otimes |g\rangle
\rightarrow |0\rangle_1 \otimes |0\rangle_{_R} \otimes
(\cos{\theta}|0\rangle_2 - \sin{\theta}|1\rangle_2)\otimes
|g\rangle. \label{transfer3}
\end{eqnarray}
If the operation time of the second step is taken as $g_2 t=3\pi/2$,
the final state after the information transfer is
\begin{eqnarray}           
|\psi\rangle_3^{transfer} = |0\rangle_1 \otimes |0\rangle_{_R}
\otimes (\cos{\theta}|0\rangle_2 + \sin{\theta}|1\rangle_2) \otimes
|g\rangle. \label{transfer4}
\end{eqnarray}
This is just the result of the quantum state transfer between the
two resonators $r_1$ and $r_2$ from the initial state
$|\psi\rangle_0^{transfer}$.

\subsection{Controlled-phase gate on $r_1$ and $r_2$}

C-phase gate is an important universal two-qubit gate. In the basis
of two resonators $|r\rangle_1$ and $|r\rangle_2$
$\{|0\rangle_1|0\rangle_2, |0\rangle_1|1\rangle_2,
|1\rangle_1|0\rangle_2, |1\rangle_1|1\rangle_2 \}$, a matrix of the
gate can be expressed as
\begin{equation}        
U^{cp}_{g,e}=\left(
\begin{array}{cccc}
1 & 0 & 0 & 0 \\
0 & -1 & 0 & 0 \\
0 & 0 & 1 & 0 \\
0 & 0 & 0 & 1
\end{array}
\right), \label{cp}
\end{equation}
which means a minus phase should be generated if and only if the two
qubits are in the state $|0\rangle_1|1\rangle_2$. In our processor,
the c-phase gate on the resonators $r_1$ and $r_2$ can be completed
with five steps by combining the resonance operations between the
quantum bus $R$ and the resonator $r_j$, and those between $R$ and
$q$ with the two transitions $|g\rangle \leftrightarrow |e\rangle$
and $|e\rangle \leftrightarrow |f\rangle$.

By taking the coupling strength between $q$ and $R$ much smaller
than the anharmonicity of $q$
($g_{g,e}\ll\omega_{g,e}-\omega_{e,f}$), the interactions between
$R$ and  $q$ with the two transitions of $|g\rangle \leftrightarrow
|e\rangle$ and $|e\rangle \leftrightarrow |f\rangle$ can be reduced
into those of  two individual two-energy-level qubits with $R$,
whose Hamiltonians are
\begin{eqnarray}        
H_{g,e}^{R,q} = g_{g,e}\left(a_{_R}\sigma_{g,e}^{+}e^{i\Delta_{g,e}
t} + a_{_R}^{+}\sigma_{g,e}^{-}e^{-i\Delta_{g,e} t}\right)
\label{Rq1}
\end{eqnarray}
and
\begin{eqnarray}        
H_{e,f}^{R,q} &=&
g_{e,f}\left(a_{_R}\sigma_{e,f}^{+}e^{i\Delta_{e,f} t} +
a_{_R}^{+}\sigma_{e,f}^{-}e^{-i\Delta_{e,f} t}\right), \label{Rq2}
\end{eqnarray}
respectively. In the condition of resonance interactions between $R$
and $q$ with the transitions $|g\rangle \leftrightarrow |e\rangle$
($\Delta_{g,e}=0$) and $|e\rangle \leftrightarrow |f\rangle$
($\Delta_{e,f}=0$), the time-evolution operation of the system
undergoing the Hamiltonians $H_{g,e}^{R,q}$ and $H_{e,f}^{R,q}$ are
\cite{Scully}
\begin{eqnarray}        
\begin{split}
U_{g,e}^{R,q} =& \exp(-i H_{g,e}^{R,q} t) \\
=& \cos(g_{g,e}t\sqrt{a^{+}a+1})|e\rangle\langle e|  + \cos(g_{g,e}t\sqrt{a^{+}a})|g\rangle \langle g|  \\
&
-i\frac{\sin(g_{g,e}t\sqrt{a^{+}a+1})}{\sqrt{a^{+}a+1}}a|e\rangle\langle
g| -
ia^{+}\frac{\sin(g_{g,e}t\sqrt{a^{+}a+1})}{\sqrt{a^{+}a+1}}|g\rangle\langle
e| \label{u1}
\end{split}
\end{eqnarray}
and
\begin{eqnarray}        
\begin{split}
U_{e,f}^{R,q} =& \exp(-i H_{e,f}^{R,q} t) \\
=& \cos(g_{e,f}t\sqrt{a^{+}a+1})|f\rangle\langle f| + \cos(g_{e,f}t\sqrt{a^{+}a})|e\rangle \langle e|  \\
&
-i\frac{\sin(g_{e,f}t\sqrt{a^{+}a+1})}{\sqrt{a^{+}a+1}}a|f\rangle\langle
e| -
ia^{+}\frac{\sin(g_{e,f}t\sqrt{a^{+}a+1})}{\sqrt{a^{+}a+1}}|e\rangle\langle
f|, \label{u2}
\end{split}
\end{eqnarray}
respectively.

Supposing the initial state of the processor is
\begin{eqnarray}        
\begin{split}
|\psi\rangle_0^{cp} \;\;=\;\;& \frac{1}{2}(\cos\theta_1  |0\rangle_1
+ \sin\theta_1 |1\rangle_1)
\otimes   |0\rangle_{_R} \otimes |g\rangle \otimes (\cos\theta_2 |0\rangle_2 + \sin\theta_2 |1\rangle_2) \\
\;\;=\;\;& \frac{1}{2}(\alpha_1 |0\rangle_1 |0\rangle_{_R} |g\rangle
|0\rangle_2 + \alpha_2 |0\rangle_1 |0\rangle_{_R} |g\rangle
|1\rangle_2 \\
&+ \alpha_3 |1\rangle_1 |0\rangle_{_R} |g\rangle |0\rangle_2 +
\alpha_4 |1\rangle_1 |0\rangle_{_R} |g\rangle |1\rangle_2).
\label{cp0}
\end{split}
\end{eqnarray}
Here, the amplitudes $\alpha_1=\cos\theta_1 \cos\theta_2$,
$\alpha_2=\cos\theta_1 \sin\theta_2$, $\alpha_3=\sin\theta_1
\cos\theta_2$, and $\alpha_4=\sin\theta_1 \sin\theta_2$. The five
steps for the construction of our c-phase gate on $r_1$ and $r_2$
can be described in detail as follows.

First, turning on the coupling strength between $R$ and $r_1$ with
$g_1=g_{g,e}$, and turning off the interaction between $R$ and
$r_2$, the state of the processor can evolve from
$|\psi\rangle_0^{cp}$ to
\begin{eqnarray}        
\begin{split}
|\psi\rangle_1^{cp} \;\;=\;\;& \frac{1}{2}(\alpha_1 |0\rangle_1
|0\rangle_{_R} |g\rangle |0\rangle_2 + \alpha_2 |0\rangle_1
|0\rangle_{_R} |g\rangle |1\rangle_2 \\
&- \alpha_3 |0\rangle_1 |0\rangle_{_R} |e\rangle |0\rangle_2 -
\alpha_4 |0\rangle_1 |0\rangle_{_R} |e\rangle |1\rangle_2)
\label{cp2}
\end{split}
\end{eqnarray}
with an operation time of $t=\pi/\sqrt{2}g_1$ \cite{Tao}.

Second, tuning the frequency of $q$ to detune with $R$ largely and
turning off the coupling between $R$ and $r_1$, one can get the
state of the processor as
\begin{eqnarray}        
\begin{split}
|\psi\rangle_2^{cp} \;\;=\;\;& \frac{1}{2}(\alpha_1 |0\rangle_1
|0\rangle_{_R} |g\rangle |0\rangle_2 - i\alpha_2 |0\rangle_1
|1\rangle_{_R} |g\rangle |0\rangle_2  \\
&- \alpha_3 |0\rangle_1 |0\rangle_{_R} |e\rangle |0\rangle_2 +
i\alpha_4 |0\rangle_1 |1\rangle_{_R} |e\rangle |0\rangle_2)
\label{cp3}
\end{split}
\end{eqnarray}
after the time of $g_2 t=\pi/2$ when the coupling between $R$ and
$r_2$ is turned on.

Third, resonating $R$ and $q$ with the transition of $|e\rangle
\leftrightarrow |f\rangle$ with a time of $g_{e,f} t=\pi$, and
keeping $R$ uncoupled to $r_1$ and $r_2$, the state of the the
processor becomes
\begin{eqnarray}        
\begin{split}
|\psi\rangle_3^{cp} \;\;=\;\;& \frac{1}{2}(\alpha_1 |0\rangle_1
|0\rangle_{_R} |g\rangle |0\rangle_2 - i\alpha_2 |0\rangle_1
|1\rangle_{_R} |g\rangle |0\rangle_2 \\
&- \alpha_3 |0\rangle_1 |0\rangle_{_R} |e\rangle |0\rangle_2 -
i\alpha_4 |0\rangle_1 |1\rangle_{_R} |e\rangle |0\rangle_2).
\label{cp4}
\end{split}
\end{eqnarray}

Fourth, repeating the second step, one can get the state of the
processor as
\begin{eqnarray}        
\begin{split}
|\psi\rangle_4^{cp} \;\;=\;\;& \frac{1}{2}(\alpha_1 |0\rangle_1
|0\rangle_{_R} |g\rangle |0\rangle_2 - \alpha_2 |0\rangle_1
|0\rangle_{_R} |g\rangle |1\rangle_2 \\
&- \alpha_3 |0\rangle_1 |0\rangle_{_R} |e\rangle |0\rangle_2 -
\alpha_4 |0\rangle_1 |0\rangle_{_R} |e\rangle
|1\rangle_2).\label{cp5}
\end{split}
\end{eqnarray}

Fifth, repeating the first step, we can get the state
\begin{eqnarray}        
\begin{split}
|\psi\rangle_f^{cp} \;\;=\;\;& \frac{1}{2}(\alpha_1 |0\rangle_1
|0\rangle_{_R} |g\rangle |0\rangle_2 - \alpha_2 |0\rangle_1
|0\rangle_{_R} |g\rangle |1\rangle_2 \\
& + \alpha_3 |1\rangle_1 |0\rangle_{_R} |g\rangle |0\rangle_2 +
\alpha_4 |1\rangle_1 |0\rangle_{_R} |g\rangle |1\rangle_2).
\label{cp7}
\end{split}
\end{eqnarray}
This is just the result of our c-phase gate  on $r_1$ and $r_2$ with
the initial state $|\psi\rangle_{0}^{cp}$.

\section{Possible experimental implementation}

Resonance operation between a superconducting qubit and a 1D
superconducting resonator has been used to achieve some basic tasks
in quantum information processing, such as generating Fock states in
a superconducting quantum circuit \cite{Hofheinz}, realizing the
NOON state entanglement on two superconducting microwave resonators
\cite{Wang}, constructing the resonant quantum gates on charge
qubits in circuit QED \cite{Haack} or on resonators \cite{Hua2}, and
completing a fast scheme to generate NOON state entanglement on two
resonators \cite{Su}. To get a high-fidelity resonant operation
between the qubit and the resonator, the magnetic flux with fast
tunability is required.

To show the performance of our schemes for quantum state transfer
and the c-phase gate, we simulate their fidelities by using the
whole Hamiltonian in each step. In our simulations,  the parameters
are chosen as: $g_1/(2\pi)$ and $g_2/(2\pi)$ can be tuned from $0$
MHz to $50$ MHz, $\omega_{_R}/(2\pi)=6.65$ GHz \cite{Peropadre},
$\delta=\omega_{g,e}/(2\pi)-\omega_{e,f}/(2\pi)=0.72$ GHz
\cite{Hoi}, $g_{g,e}/(2\pi)=g_{e,f}/(2\sqrt{2}\pi)=13$ MHz,
$\kappa_1^{-1}=\kappa_2^{-1}=\kappa_R^{-1} \equiv \kappa^{-1}=50$
$\mu$s, and
$\gamma_{g,e}^{-1}=\frac{1}{2}\gamma_{e,f}^{-1}=\gamma_{\phi,e}^{-1}=\gamma_{\phi,f}^{-1}
\equiv \Gamma^{-1}=50$ $\mu$s. The transition frequency of a
transmon qutrit can be tune with a range of about $2.5$ GHz
\cite{Schreier}, which is enough for us to make it detune with $R$
largely. The maximal values of $g_1/(2\pi)$ and $g_2/(2\pi)$ taken
by us are $50$ MHz as the rotation-wave approximation can work well
when the coupling strength is much smaller than the frequency of $R$
and a theoretic predict of the coupling strength between two
superconducting resonators can reach $1.2$ GHz \cite{Peropadre}.

The process for the generation of the initial states of
$|\psi\rangle_0^{transfer}$ and $|\psi\rangle_0^{cp}$ are not
included in our simulations. To prepare the initial states, one
should perform a proper single-qubit operation on $q$ and send the
information from $q$ to $r_j$ by using the resonance operation, the
same as the one in the first step for the construction of our
c-phase gate. Here, the interactions which do not attend the
resonance operation should be turned off. The single-qubit operation
on a superconducting qubit has been realized in experiment with a
quantum error smaller than $0.0006$ \cite{Barends}, which has little
influence on our schemes. By taking the energy relaxation rate of
the qutrit, the decay rates of resonators, and $g_{g,e}$ and $g_j$
into account, the generation of the initial states just increases a
little error value of the fidelities of the quantum state transfer
and the c-phase gate.

\subsection{Fidelity for our quantum state transfer}

We numerically simulate the populations (vary with the operation
time) of a microwave photon in $r_1$, $R$, and $r_2$, shown in Fig.
2. The definition of the population is
\begin{eqnarray}        
P_m &=& \langle \psi_m|\rho(t)|\psi_m\rangle. \label{p}
\end{eqnarray}
Here $m=1$, $2$, $3$. $|\psi_1\rangle=|1\rangle_1 |0\rangle_{_R}
|g\rangle |0\rangle_2$, $|\psi_2\rangle=|0\rangle_1 |1\rangle_{_R}
|g\rangle |0\rangle_2$, and $|\psi_3\rangle=|0\rangle_1
|0\rangle_{_R} |g\rangle |1\rangle_2$. $\rho(t)$ is the realistic
density operator of the processor for the quantum state transfer
from the initial state $|1\rangle_1 |0\rangle_{_R} |g\rangle
|0\rangle_2$. The parameters taken in the first step in our scheme
are: $\omega_{g,e}/(2\pi)=5$ GHz, $g_1/(2\pi)=50$ MHz,
$g_2/(2\pi)=0$ MHz. In the second step, the parameters are:
$g_1/(2\pi)=0$ MHz, $g_2/(2\pi)=50$ MHz, and the other parameters
are the same as the ones in the first step.

From the numerical simulation, the quantum state transfer between
$r_1$ and $r_2$ with $\theta=\frac{\pi}{4}$ can reach a fidelity of
$99.97\%$ within $10$ ns by using the definition of the fidelity as
$F=\langle \Psi|\rho(t)|\Psi\rangle$
($|\Psi\rangle=|\psi\rangle_2^{transfer}$) with the initial state
$|\psi\rangle_0^{transfer}$.  In the inset in  Fig. 2, we give the
three conditions of the populations with different decay rates of
$r_1$, $r_2$, and $R$.

\begin{figure}[tpb!]               
\begin{center}
\includegraphics[width=8.5 cm,angle=0]{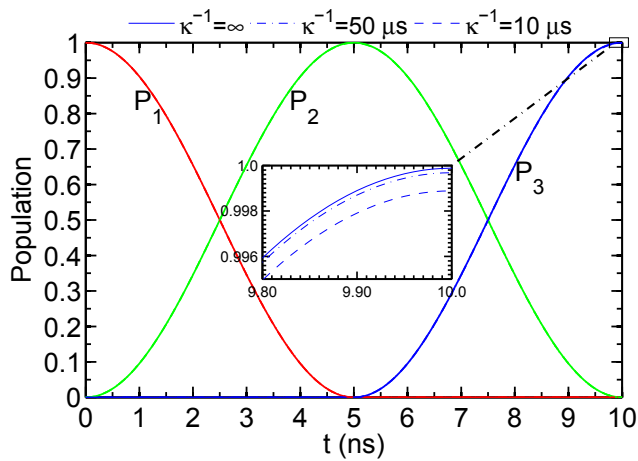}
\end{center}
\caption{(Color online)  The populations of a microwave photon in
$r_1$, $R$, and $r_2$. $P_1$, $P_2$, and $P_3$  with the red, green,
and blue solid lines represent the populations of the  microwave
photon in $r_1$, $R$, and $r_2$, respectively. The inset shows the
populations varying with the decay rates of the resonators, in which
the solid, the dot dash, and the dotted lines represent those with
the decay rates of the resonators $\kappa^{-1}=\infty$ $\mu$s,
$\kappa^{-1}=50$ $\mu$s, and $\kappa^{-1}=10$ $\mu$s, respectively.
} \label{fig2}
\end{figure}


\begin{figure}[tpb!]               
\begin{center}
\includegraphics[width=16 cm,angle=0]{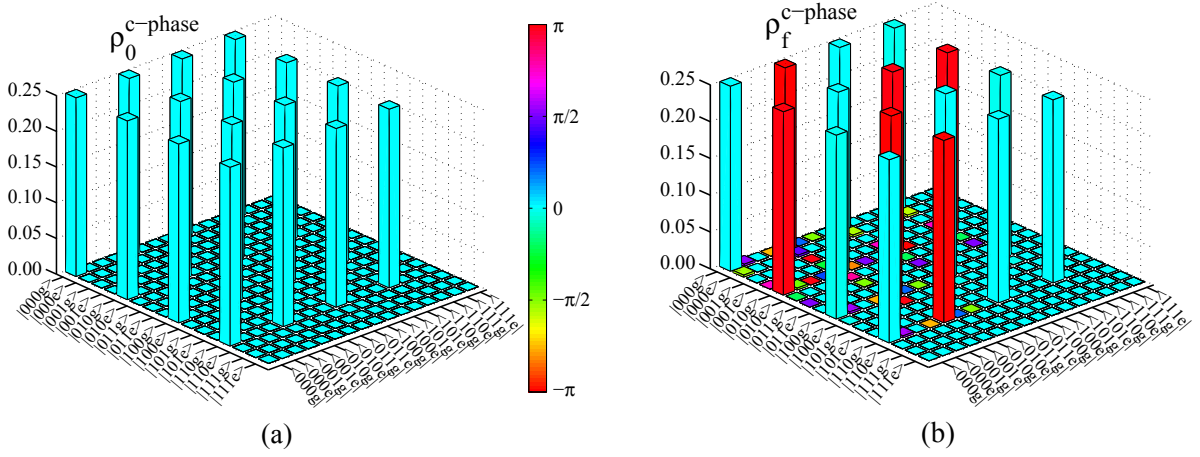}
\end{center}
\caption{(Color online) (a) The density operator $\rho_0$ of the
initial state $|\psi\rangle_0^{cp}$ of our processor. (b) The
realistic density operator $\rho_f^{c-phase}$ of the final state
$\left\vert \psi_{f}\right\rangle$ after our c-phase gate operation
is performed on the two microwave-photon resonators. The color bar
indicates the phase information of the density matrix elements. }
\label{fig3}
\end{figure}

\subsection{Fidelity for our c-phase gate}

We calculate the fidelity of our c-phase gate by using the
average-gate-fidelity definition \cite{Zhang,Johansson}
\begin{eqnarray}          
F = \left(\frac{1}{2\pi}\right)^2 \int_0^{2\pi} \int_0^{2\pi}
\langle \Psi_{ideal}|\rho(t)|\Psi_{ideal}\rangle d\theta_1
d\theta_2. \label{fidelity}
\end{eqnarray}
Here, $|\Psi_{ideal}\rangle$ is the final state
($|\psi\rangle_f^{cp}$) of the processor by using the ideal c-phase
gate operation on the initial state $|\psi_0^{cp}\rangle$. $\rho(t)$
is the realistic density operator after our c-phase gate operation
on the initial state with the Hamiltonian $H$. The fidelity of our
c-phase gate reaches $99.66\%$ within $91.5$ ns by using the
parameters taken in each step as shown in Table 1. Here, if we take
$\theta_1=\theta_2=\pi/4$ in Eq. (\ref{cp0}) as an example, the
density operators of $|\psi\rangle_0^{cp}$ and the real final state
are shown in Fig. 3 (a) and (b), respectively.

\begin{table}
\centering \caption{Parameters for the construction of the c-phase
gate on $r_1$ and $r_2$. }
\begin{tabular}{lllllllllllll}
\hline\hline
\multirow{2}{*}{Step}   &  &  &  & $g_1/(2\pi)$        &  &  &  & $g_2/(2\pi)$       &  &   &  & $\omega_{g,e}/(2\pi)$ \\
                        &  & &  & \;\;{\footnotesize (MHz)}   &  & &  &  \;\;{\footnotesize (MHz)}
                       &  &  &  &  \;\;{\footnotesize (GHz)}\\ \hline
\;\;i   &  & &  & \;\;\;\;\;$13$     &  & &  & \;\;\;\;\;\;$0$    &  & &  & \;\;\;$6.65$ \\ 
\;\;ii   &  & &  & \;\;\;\;\;\;$0$     &  & &  & \;\;\;\;\;$50$     &  & &  & \;\;\;\;\;$5$ \\ 
\;\;iii  &  & &  & \;\;\;\;\;\;$0$    &  & &  & \;\;\;\;\;\;$0$     &  & &  & \;\;\;$7.37$ \\ 
\;\;iv   &  & &  & \;\;\;\;\;\;$0$    &  &  &  & \;\;\;\;\;$50$     &  & &  & \;\;\;\;\;$5$ \\ 
\;\;v  &  & &  & \;\;\;\;\;$13$      &  & &  & \;\;\;\;\;\;$0$     &
& &  & \;\;\;$6.65$ \\ \hline\hline
\end{tabular}\label{table1}
\end{table}

Actually, the fidelity of our c-phase gate is influenced by the
decay rates $\kappa$ of the resonators, the energy relaxation rate
$\Gamma$ of $q$, and the anharmonicity $\delta$ of $q$, shown in
Fig. 4. In Fig. 4(a), we show the fidelity of the gate varying with
the decay rates and the energy relaxation rate of the resonators and
$q$ ($\kappa=\Gamma$). The fidelity of the gate is numerically
simulated by using different optimal parameters corresponding to
different $\Gamma$ (keeping $\delta=0.72$ GHz unchanged) as the
competition between the operation time (leads to the error from the
coherence time of the qutrit) and the coupling strength between the
qutrit and the bus $R$ (leads to the error from the anharmonicity of
the qutrit). Here, in order to choose $\Gamma^{-1}=10$, $20$, $30$,
$40$, and $50$ $\mu$s, we take
$g_{g,e}/(2\pi)=g_{e,f}/(2\sqrt{2}\pi)=22$, $19$, $13$, $13$, and
$13$ MHz, respectively. The corresponding operation times are
$t=58.1$, $65.8$, $91.5$, $91.5$, and $91.5$ ns, respectively. By
using $\kappa=\omega_r/Q$ ($\omega_r$ is the frequency of the
resonator) \cite{Blais}, $\kappa^{-1}=10$ $\mu$s  corresponds to a
quality factor $Q\sim 4.2\times 10^{5}$ of the resonators. In Fig.
4(b), the anharmonicity of the qutrit influences the fidelity with a
small value as the coupling strength $g_{g,e}$ is much smaller than
$\delta$, which means that the transmon qutrit in our processor does
not require a large anharmonicity.

\begin{figure}[tpb!]               
\begin{center}
\includegraphics[width=10.0 cm,angle=0]{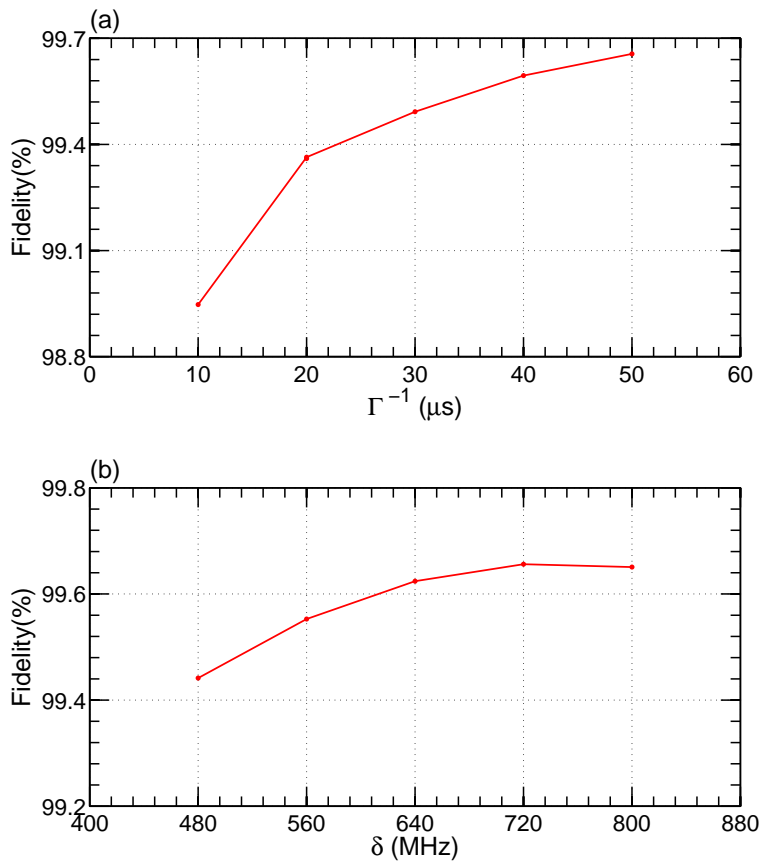}
\end{center}
\caption{(Color online) The fidelity of our c-phase gate on the two
microwave-photon resonators  $r_1$ and $r_2$ which varies with the
parameters $\kappa^{-1}=\Gamma^{-1}$ (a) and $\delta$ (b),
respectively. } \label{fig4}
\end{figure}

\begin{table}
\centering \caption{Parameters for the construction of the c-phase
gate on $r_1$ and $r_2$. }
\begin{tabular}{lllllllllllll}
\hline\hline
\multirow{2}{*}{Step}   &  &  &  & $g_1/(2\pi)$        &  &  &  & $g_2/(2\pi)$       &  &   &  & $\omega_{g,e}/(2\pi)$ \\
                        &  & &  & \;\;{\footnotesize (MHz)}   &  & &  &  \;\;{\footnotesize (MHz)}
                       &  &  &  &  \;\;{\footnotesize (GHz)}\\ \hline
\;\;i   &  & &  & \;\;\;\;\;$50$     &  & &  & \;\;\;\;\;\;$0$    &  & &  & \;\;\;\;\;\;$5$ \\ 
\;\;ii   &  & &  & \;\;\;\;\;\;$0$     &  & &  & \;\;\;\;\;\;$0$     &  & &  & \;\;\;\;$6.65$ \\ 
\;\;iii  &  & &  & \;\;\;\;\;\;$0$    &  & &  & \;\;\;\;\;$50$     &  & &  & \;\;\;\;\;\;$5$ \\ 
\;\;iv   &  & &  & \;\;\;\;\;\;$0$    &  &  &  & \;\;\;\;\;\;$0$     &  & &  & \;\;\;\;$7.37$ \\ 
\;\;v    &  & &  & \;\;\;\;\;\;$0$    &  & &  & \;\;\;\;\;$50$     &  & &  & \;\;\;\;\;\;$5$ \\ 
\;\;vi   &  & &  & \;\;\;\;\;\;$0$     &  & &  & \;\;\;\;\;\;$0$     &  & &  & \;\;\;\;$6.65$ \\ 
\;\;vii  &  & &  & \;\;\;\;\;$50$      &  & &  & \;\;\;\;\;\;$0$     &  & &  & \;\;\;\;\;\;$5$ \\ \hline\hline
\end{tabular}\label{table}
\end{table}

Actually, the fidelity of our c-phase gate is influenced by the
decay rates $\kappa$ of the resonators,  the anharmonicity $\delta$
of $q$, and the energy relaxation rate $\Gamma$ of $q$, shown in
Fig.\ref{fig4}. Here, the anharmonicity of the qutrit influences the
fidelity with a small value as the coupling strength $g_{g,e}$ is
much smaller than $\delta$, shown in Fig.\ref{fig4}(b), which means
that the transmon qutrit in our processor does not require a large
anharmonicity.

\section{Conclusion}

To show our processor can be used for an effective quantum
computation based on resonators, we have given the scheme to achieve
the quantum state transfer between two resonators and the one for
the c-phase gate on them. These two schemes are just based on the
Fock states $|0\rangle_j$ and $|1\rangle_j$ of the resonators $r_j$.
The fidelities of our quantum state transfer and  c-phase gate reach
$99.97\%$ and $99.66\%$ within  $10$ ns and $91.5$ ns, respectively.
In our processor, a single-qubit operation on the resonator $r_j$
can be achieved with the following steps: 1), one should transfer
the information from  $r_j$ to the qutrit with the resonance
operation between them. 2), one can take the single-qubit gate on
the qutrit. 3), one should transfer the information from the qutrit
to $r_j$. It is worth noticing that there are two steps with
resonance operations in our scheme for the single-qubit operation on
a microwave-photon resonator. Each resonance operation can generate
a $-$ phase for the state $|1\rangle_j$ of the resonator $r_j$ or
the state $|e\rangle$ of the qutrit. The two steps with resonance
operations can just eliminate this unwanted phase generated by each
resonance operation as $(-1)^2=1$. So, the single-qubit operation on
the qutrit is convenient without considering the additional phase
generated by the resonance operations. To readout the information of
the photon states in $r_j$, one can also transfer the information of
the photon from $r_j$ (based on the Fock states $|0\rangle_j$ and
$|1\rangle_j$) to the qutrit (based on the states $|g\rangle$ and
$|e\rangle$) and then readout the state of the qutrit. To achieve
the quantum non-demolition detection on the resonator $r_j$, one can
use a low-quality resonator coupled to the qutrit $q$ to detect the
information in the quantum bus R \cite{Johnson} which comes from
$r_j$. By using the resonators which can catch and release the
microwave photons \cite{Yin}, our processor maybe play an important
role in quantum communication.

In summary, we have proposed a quantum processor composed of some 1D
superconducting resonators $r_j$ (quantum information carriers)
which are coupled to a common 1D superconducting resonator $R$ (the
quantum bus), not the superconducting transmon qutrit, which makes
it have the capability of integrating some distant resonators for
quantum information processing on microwave photons assisted by
circuit QED. With this processor, we have presented a  scheme for
the high-fidelity state transfer between two resonators. Also, we
have given a scheme for the c-phase gate on two resonators with the
resonance operations.  With feasible parameters in experiment, the
fidelities of our two schemes are $99.97\%$ and $99.66\%$,
respectively. Maybe this processor can play an important role in
quantum communication in future.

\section{Methods}

{\bf Interaction between a resonator and a qubit.} In the
interaction picture, the Hamiltonian of a system composed of a
two-energy-level qubit coupled to a resonator (Q-R system) can be
written as (under the rotating-wave approximation):
\begin{eqnarray}           
H_I=g\left(\sigma^{+}ae^{i\Delta t}+\sigma^{-}a^{+}e^{-i\Delta
t}\right). \label{method}
\end{eqnarray}
Here, $g$ is the coupling strength between the qubit and the
resonator. $\sigma^{+}=|e\rangle \langle g|$ and $a^{+}$ are the
create operators of the qubit and the resonator, respectively.
$\Delta=\omega_{q}-\omega_{r}$. $\omega_{q}$ ($\omega_{r}$) is the
transition frequency of the qubit (resonator).

The state ($|\psi\rangle$) of the Q-R system can be solved with the
equation of motion
\begin{eqnarray}           
i \frac{\partial|\psi(t)\rangle}{\partial t}= H_{I} |\psi(t)\rangle,
\label{1}
\end{eqnarray}
in which $|\psi(t)\rangle$ is a linear combination of the states
$|e\rangle |n\rangle_r$ and $|g\rangle |n\rangle_r$, that is,
\begin{eqnarray}           
|\psi(t)\rangle=\sum_{n}\left[ c_{e,n}(t)|e\rangle
|n\rangle_r+c_{g,n}(t)|g\rangle |n\rangle_r\right]. \label{2}
\end{eqnarray}
Here, $c_{e,n}(t)$ and $c_{g,n}(t)$ are the slowly varying
probability amplitudes. $|n\rangle_r$ is the Fock state of the
resonator. Because  the only transitions between $|e\rangle
|n\rangle_r$ and $|g\rangle |n+1\rangle_r$ can be caused by the
Hamiltonian $H_I$, we just need to consider the evolutions of
$c_{e,n}(t)$ and $c_{g,n+1}(t)$.

By combining Eqs.(\ref{1}) and  (\ref{2}), one can get
\begin{eqnarray}            
\begin{split}
\frac{\partial c_{e,n}}{\partial t} =& -i g_j c_{g,n+1}e^{i\Delta t}, \\
\frac{\partial c_{g,n+1}}{\partial t} =& -i g_j c_{e,n}e^{-i\Delta
t}. \label{3}
\end{split}
\end{eqnarray}
A general solution for these amplitudes is
\begin{eqnarray}           
\begin{split}
c_{e,n}(t) =&\left\{c_{e,n}(0)\left[\cos (\frac{\Omega
t}{2})-\frac{i\Delta }{\Omega } \sin (\frac{\Omega t}{2})\right] -
\frac{2ig}{\Omega }c_{g,n+1}(0)\sin (\frac{\Omega t
}{2})\right\}e^{i\Delta t/2}, \\
c_{g,n+1}(t) =&\left\{c_{g,n+1}(0)\left[\cos (\frac{\Omega
t}{2})+\frac{i\Delta }{\Omega } \sin (\frac{\Omega t}{2})\right] -
\frac{2ig}{\Omega }c_{e,n}(0)\sin (\frac{\Omega t
}{2})\right\}e^{-i\Delta t/2}. \label{4}
\end{split}
\end{eqnarray}
Here $\Omega ^{2}=4g^{2}(n+1)+\Delta^{2}$.

\section*{ACKNOWLEDGMENTS}

The simulations were coded in PYTHON using the QUTIP library. This
work is supported by the China Postdoctoral Science Foundation under
Grant No. 2015M581061, the National Natural Science Foundation of
China under Grants No. 11474026, and the Fundamental Research Funds
for the Central Universities under Grant No. 2015KJJCA01.

\section*{Author contributions}

M.H. and M.J. completed the calculation and prepared the figures.
M.H. and F.G.  wrote the main manuscript text.  F.G. supervised the
whole project. All authors reviewed the manuscript.

\end{document}